\documentclass{emulateapj}

\usepackage{graphicx}
\usepackage[]{natbib}
\usepackage{placeins}
\usepackage{lineno}
\usepackage{ulem}

\newcommand{\degree}{\ensuremath{^\circ}}

\newcommand{\ltsima} {$\; \buildrel < \over \sim \;$}
\newcommand{\gtsima} {$\; \buildrel > \over \sim \;$}
\newcommand{\lta} {\lower.5ex\hbox{\ltsima}}
\newcommand{\gta} {\lower.5ex\hbox{\gtsima}}

\shorttitle{Detection of a thermal spectral component in the prompt emission of GRB~100724B.}
\shortauthors{Guiriec et al.}

\begin{document}
%\maketitle

%\submitted{your text}

\submitted{Accepted for publication in the Astrophysical Journal Letters November, 23 2010 (Submitted October, 20 2010)}
\title{Detection of a thermal spectral component in the prompt emission of GRB~100724B.}
%\linenumbers
%% Notice placement of commas and superscripts and use of &
%% in the author list

\author{Sylvain Guiriec\altaffilmark{1}, Valerie Connaughton\altaffilmark{1}, Michael S. Briggs\altaffilmark{1}, Michael Burgess\altaffilmark{1}, Felix Ryde\altaffilmark{2,3}, Fr\'ed\'eric Daigne\altaffilmark{4,5}, Peter M{\'e}sz{\'a}ros\altaffilmark{6}, Adam Goldstein\altaffilmark{1}, Julie McEnery\altaffilmark{7,8}, Nicola Omodei\altaffilmark{9}, P.N. Bhat\altaffilmark{1}, Elisabetta Bissaldi\altaffilmark{13}, Ascensi\'on Camero-Arranz\altaffilmark{14}, Vandiver Chaplin\altaffilmark{1}, Roland Diehl\altaffilmark{13}, Gerald Fishman\altaffilmark{10}, Suzanne Foley\altaffilmark{13}, Melissa Gibby\altaffilmark{15}, Misty M. Giles\altaffilmark{15}, Jochen Greiner\altaffilmark{13}, David Gruber\altaffilmark{13}, Andreas von Kienlin\altaffilmark{13}, Marc Kippen\altaffilmark{16}, Chryssa Kouveliotou\altaffilmark{10}, Sheila McBreen\altaffilmark{17}, Charles A. Meegan\altaffilmark{12},William Paciesas\altaffilmark{1}, Robert Preece\altaffilmark{1}, Arne Rau\altaffilmark{13}, Dave Tierney\altaffilmark{17}, Alexander J. van der Horst\altaffilmark{11} \& Colleen Wilson-Hodge\altaffilmark{10}}

\altaffiltext{1}{University of Alabama in Huntsville, NSSTC, 320 Sparkman Drive, Huntsville, AL 35805, USA}
\altaffiltext{2}{Department of Physics, Royal Institute of Technology, AlbaNova, SE-106 91 Stockholm, Sweden}
\altaffiltext{3}{The Oskar Klein Centre for Cosmo Particle Physics, AlbaNova, SE-106 91 Stockholm, Sweden}
\altaffiltext{4}{Institut d’Astrophysique de Paris – UMR 7095 Universit\'e ́Pierre et Marie Curie-Paris 06; CNRS 98 bis bd Arago, 75014 Paris, France}
\altaffiltext{5}{Institut Universitaire de France}
\altaffiltext{6}{Dept. of Astronomy \& Astrophysics, Dept. of Physics and Center for Particle Astrophysics, Pennsylvania State University, University Park, PA 16802, USA}
\altaffiltext{7}{NASA Goddard Space Flight Center, Greenbelt, MD 20771, USA}
\altaffiltext{8}{Department of Physics and Department of Astronomy, University of Maryland, College Park, MD 20742, USA}
\altaffiltext{9}{Istituto Nazionale di Fisica Nucleare, Sezione di Pisa, I-56127 Pisa, Italy}
\altaffiltext{10}{Space Science Office, VP62, NASA/Marshall Space Flight Center, Huntsville, AL 35812, USA}
\altaffiltext{11}{NASA Postdoctoral Program Fellow, NASA/Marshall Space Flight Center, 320 Sparkman Drive, Huntsville, AL 35805}
\altaffiltext{12}{Universities Space Research Association, NSSTC, 320 Sparkman Drive, Huntsville, AL 35805, USA}
\altaffiltext{13}{Max-Planck-Institut f\"ur extraterrestrische Physik (Giessenbachstrasse 1, 85748 Garching, Germany)}
\altaffiltext{14}{National Space Science and Technology Center, 320 Sparkman Drive, Huntsville, AL 35805}
\altaffiltext{15}{Jacobs Technology, Inc., Huntsville, Alabama}
\altaffiltext{16}{Los Alamos National Laboratory, PO Box 1663, Los Alamos, NM 87545, USA}
\altaffiltext{17}{University College, Dublin, Belfield, Stillorgan Road, Dublin 4, Ireland}

\email{sylvain.guiriec@nasa.gov, sylvain.guiriec@uah.edu}

\begin{abstract}
Observations of GRB~{\it 100724B} with the Fermi Gamma-Ray Burst Monitor (GBM)
find that the spectrum is dominated by the typical Band functional form,
which is usually taken to represent a non-thermal emission component, but also includes a statistically highly significant thermal spectral contribution.
The simultaneous observation of the thermal and non-thermal components
allows us to confidently identify the two emission components.
The fact that these seem
to vary independently favors the idea that the thermal component is of photospheric origin
while the dominant non-thermal emission occurs at larger radii.
Our results imply either
a very high efficiency for the non-thermal process, or a very small size of the region at the
base of the flow, both quite challenging for the standard fireball model.
These problems are resolved if the jet is initially highly magnetized and has a substantial Poynting flux.
\end{abstract}


%%\keywords{globular clusters: general --- globular clusters: individual(NGC 6397,
%%NGC 6624, NGC 7078, Terzan 8}




\section{Introduction}
The prompt emission detected from Gamma-Ray Bursts (GRBs)
 is believed to originate at large distances from the central engine, from within an ultrarelativistic outflow \citep{Piran:2004}. 
This ultra-relativistic motion is necessary to avoid strong $\gamma\gamma$ annihilation, a signature that is not observed
 \citep[see, e.g., ][]{Piran:1999}.
Thermal emission is naturally expected in such a scenario.
Indeed, since the densities at the base of the relativistic flow are very large, 
the medium is optically thick to radiation owing to Thomson scattering by entrained electrons. 
The optical depth decreases during the relativistic expansion and the outflow eventually becomes transparent for its own radiation, 
at the photospheric radius. Any internal energy that is still carried out by the flow can be radiated at the photosphere and
 will be observed as a thermal component in the prompt spectrum. This expected
photospheric emission in GRB spectra was early suggested on such theoretical grounds by~\citet{Goodman:1986}, \cite{Meszaros:2002}, and \citet{Rees:2005}, among others.  
The non-thermal component observed in the spectrum has to be produced by another mechanism in the optically thin region, 
i.e., well above the photosphere. Due to the ultra-relativistic motion, this difference in the radius of the emission implies a delay between the observation of the two components that is usually small compared to the typical duration of a long GRB, 
and is also small compared to the typical duration of time intervals used for time-dependent spectroscopic analysis. 
The thermal and non-thermal components should then appear superimposed for the observer~\citep[e.g.][]{Meszaros:2000}.
\citet{Daigne:2002} pointed out that in the standard fireball model, 
the photospheric component can easily be dominant in the spectrum if the efficiency $f_\mathrm{NT}$ of the mechanism responsible for the non-thermal emission is only moderate ($f_\mathrm{NT}\lta 40\, \%$).

Observationally,~\citet{Ghirlanda:2003}, \citet{Ryde:2004,Ryde:2005}, and \citet{Ryde:2010} argued that a photospheric component is present in CGRO BATSE data. The limited energy range provided by BATSE (20-2000 keV), however, hampered the possibility of unambiguously identifying the emission process. Since the launch of Fermi in 2008, the combination of the Gamma Ray Burst Monitor (GBM) and the Large Area Telescope (LAT) provides an unprecedented energy range for GRB spectroscopy, and the identification of the emission processes responsible for the gamma-ray prompt emission may become a reality. GBM alone covers a wider energy range than its predecessor BATSE, and the design of its data enables finer resolution spectroscopy. This allows better constraints on spectral fits, with increasingly complex models.

GRB energy spectra in the keV-MeV energy range are usually well represented
by the Band function~\citep{Band:1993}, two power-laws, smoothly joined and parameterized
by E$_{\rm peak}$ which represents the energy at which peak power is radiated~\citep{Gehrels:1997}.
The value of the low-energy power-law index, $\alpha$, is higher than the value of the high-energy power-law index,
$\beta$, and the parameter E$_{\rm peak}$ of the Band function for GRBs generally appears to
follow predictable trends with time and flux level~\citep{Ford:1995,Guiriec:2010}. It is, however, an
empirical function rather than a physically-motivated model. The meaning of
the parameters in the context of emission and transport mechanisms taking
place in GRBs is not well understood, but is generally believed to represent the
non-thermal emission from accelerated charged particles.

In section 1, we describe our observational results consisting of a GRB prompt-emission spectrum best fit with the combination of a thermal component and a standard Band function. In section 2, we use these results to constrain the origin of the energy released in the
GRB jet.

\section{Observations}
\begin{figure}[h!]
%\begin{center}
\hspace{-0.9cm}
\includegraphics[width=0.42\textheight]{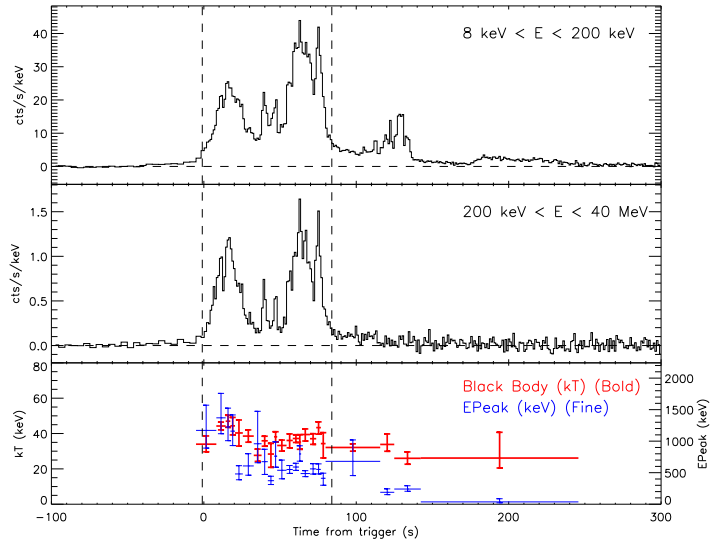}
%\end{center}
\caption{The top two panels show the signal count rates as a function of
time, as measured by the Fermi GBM detectors, from 8 to 200 keV in NaI (top) and from 200 keV to 40 MeV in BGO (middle).
The bottom panel shows the
evolution of the Band function E$_{\rm peak}$~(in blue) and the BB temperature kT~(in red) over the duration of
the burst. The vertical dashed lines indicate the period used in the time-integrated analysis.
\label{fig:figure1}}
\end{figure}

GBM is composed of 12 sodium iodide (NaI) detectors covering an energy range from 8 keV to 1 MeV and two bismuth germanate (BGO) 
detectors sensitive between 200 keV and 40 MeV~\citep{Meegan:2009}.
The instrument triggered on 24 July 2010, at T$_\mathrm{0}$=00:42:05.992 UT on the very
bright GRB~{\it 100724}B~\citep{GRB100724B_GCN_GBM}. The event was also seen at higher energies in the Fermi Large
Area Telescope (LAT)~\citep{GRB100724B_GCN_LAT}. The most precise position for the
direction of the burst is the intersection of the InterPlanetary Network
annulus obtained using GBM, Konus-WIND~\citep{GRB100724B_GCN_Konus}, and MESSENGER data with the 90\% LAT confidence level location 
error box, and is a strip of sky centered on RA = 118.8\degree~and Dec = 75.8\degree~which is 1.2\degree~long and 0.2\degree~wide 
(K.Hurley and V.Pal'shin, private communication). Figure~\ref{fig:figure1} (top two panels) shows the
GBM light curve of GRB~{\it 100724}B in two energy bands. Multiple peaks of varying intensity are superimposed on a pre-trigger plateau,
with a decaying tail that is detected over 200 s from T$_{0}$.

We simultaneously fit the spectral data of the NaI detectors with a source angle less than 60\degree~(NaIs 0, 1, 2, 3, and 5) and the 
data of the brightest BGO detector (BGO 0) using the analysis package Rmfit 3.3rc8.
An effective area correction is applied between each of the NaIs and BGO 0 during the fit process. This
correction is used to handle possible discrepancies between the flux in the detectors due to the choice
 of the model to generate the instrument responses for instance.

\begin{figure*}[h!]
\begin{center}

%\hspace{-1.5cm}
%\hspace{-1.3cm}
\includegraphics[totalheight=0.50\textwidth,clip,angle=90]{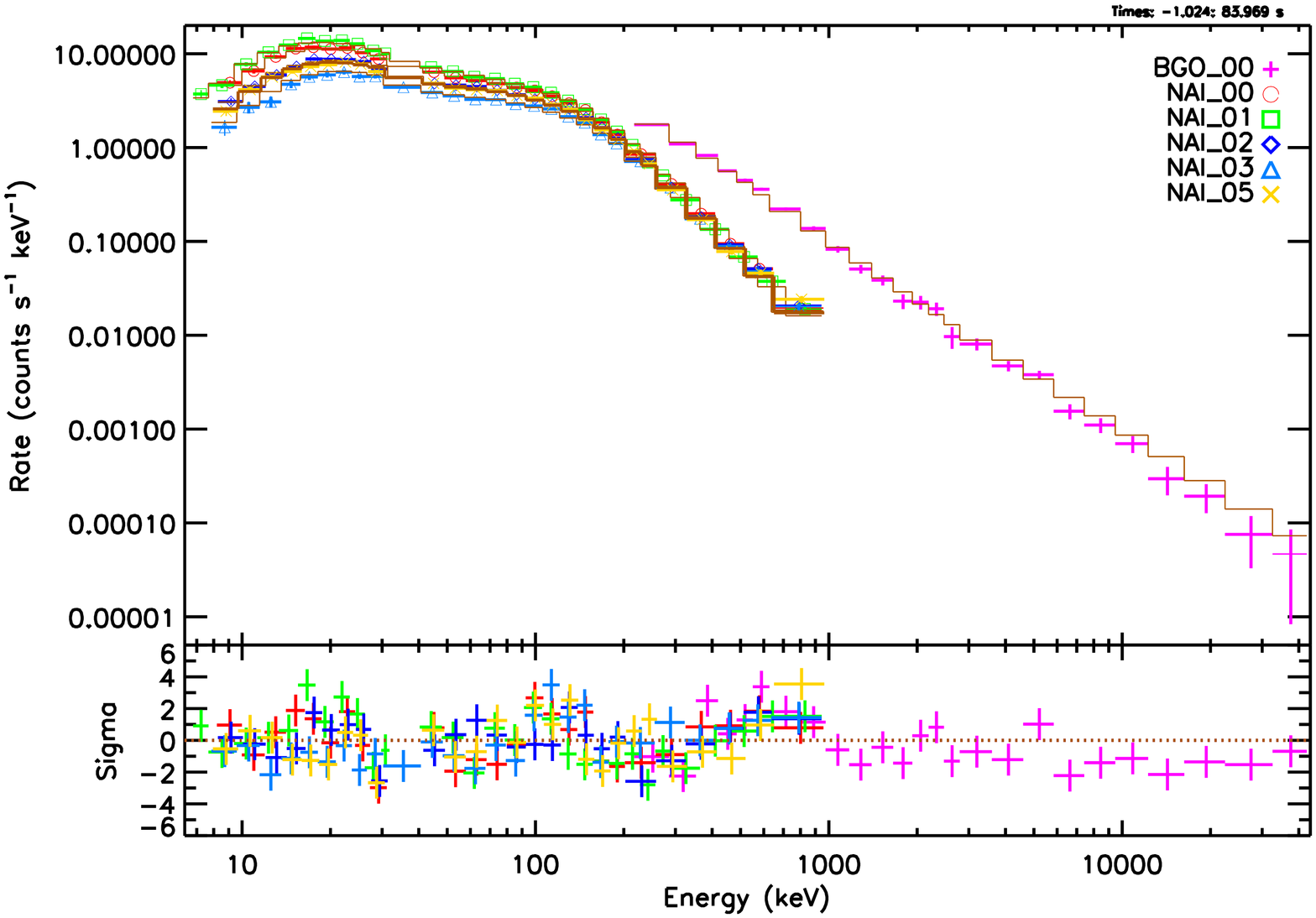}
\hspace{-0.4cm}
\includegraphics[totalheight=0.50\textwidth,clip,angle=90]{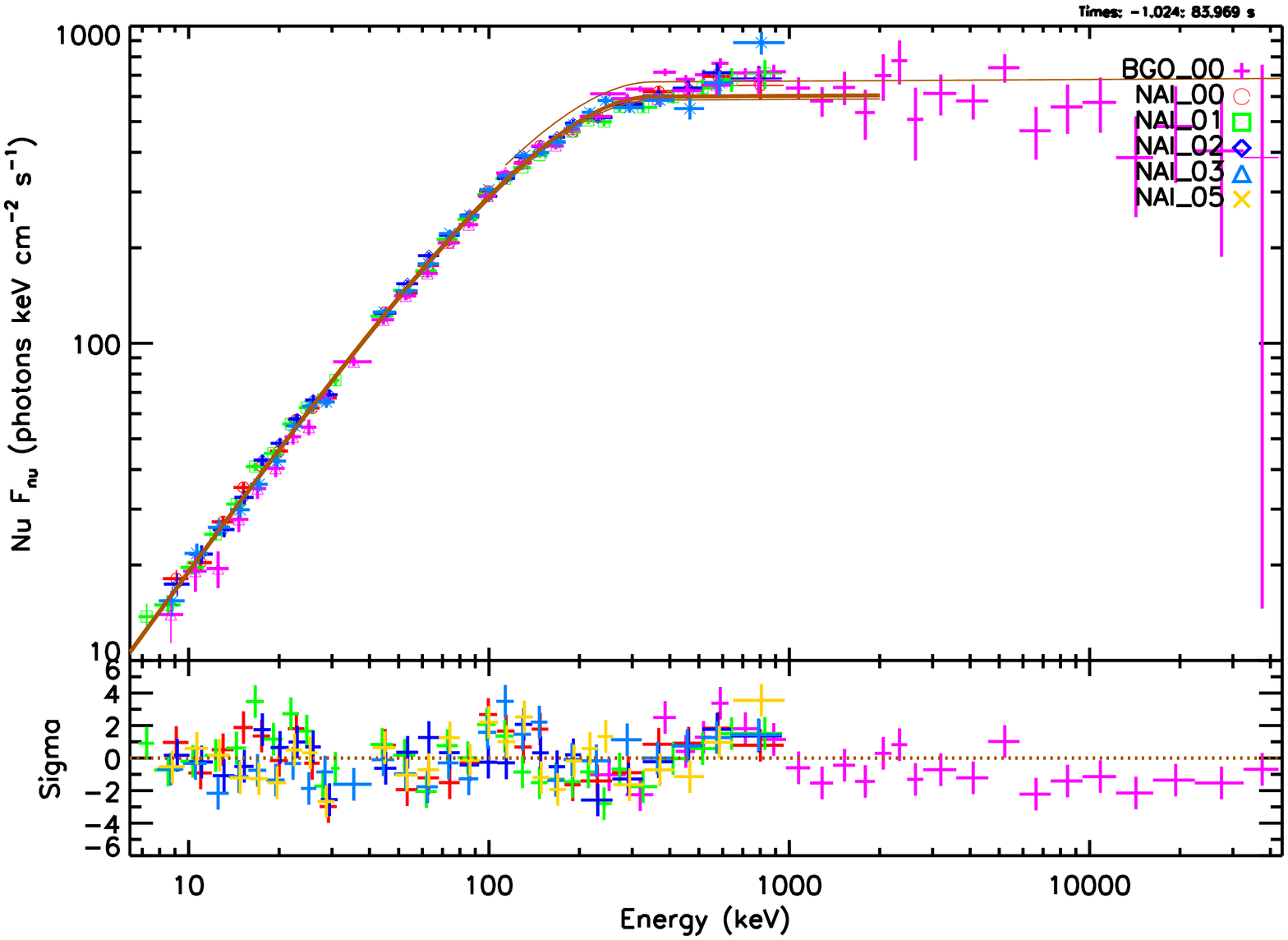}

%\hspace{-1.5cm}
%\hspace{-1.3cm}
\includegraphics[totalheight=0.50\textwidth,clip,angle=90]{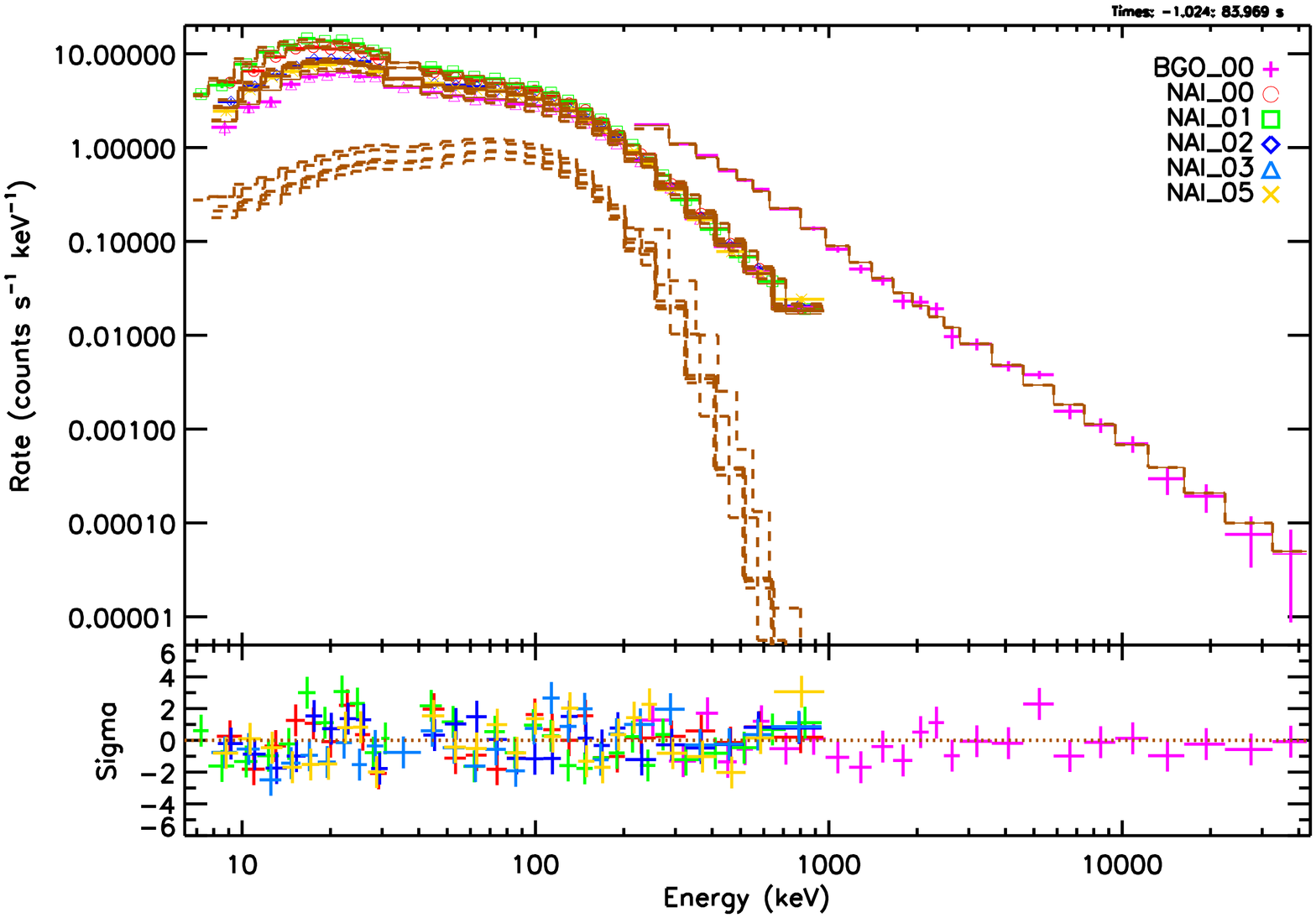}
\hspace{-0.4cm}
\includegraphics[totalheight=0.50\textwidth,clip,angle=90]{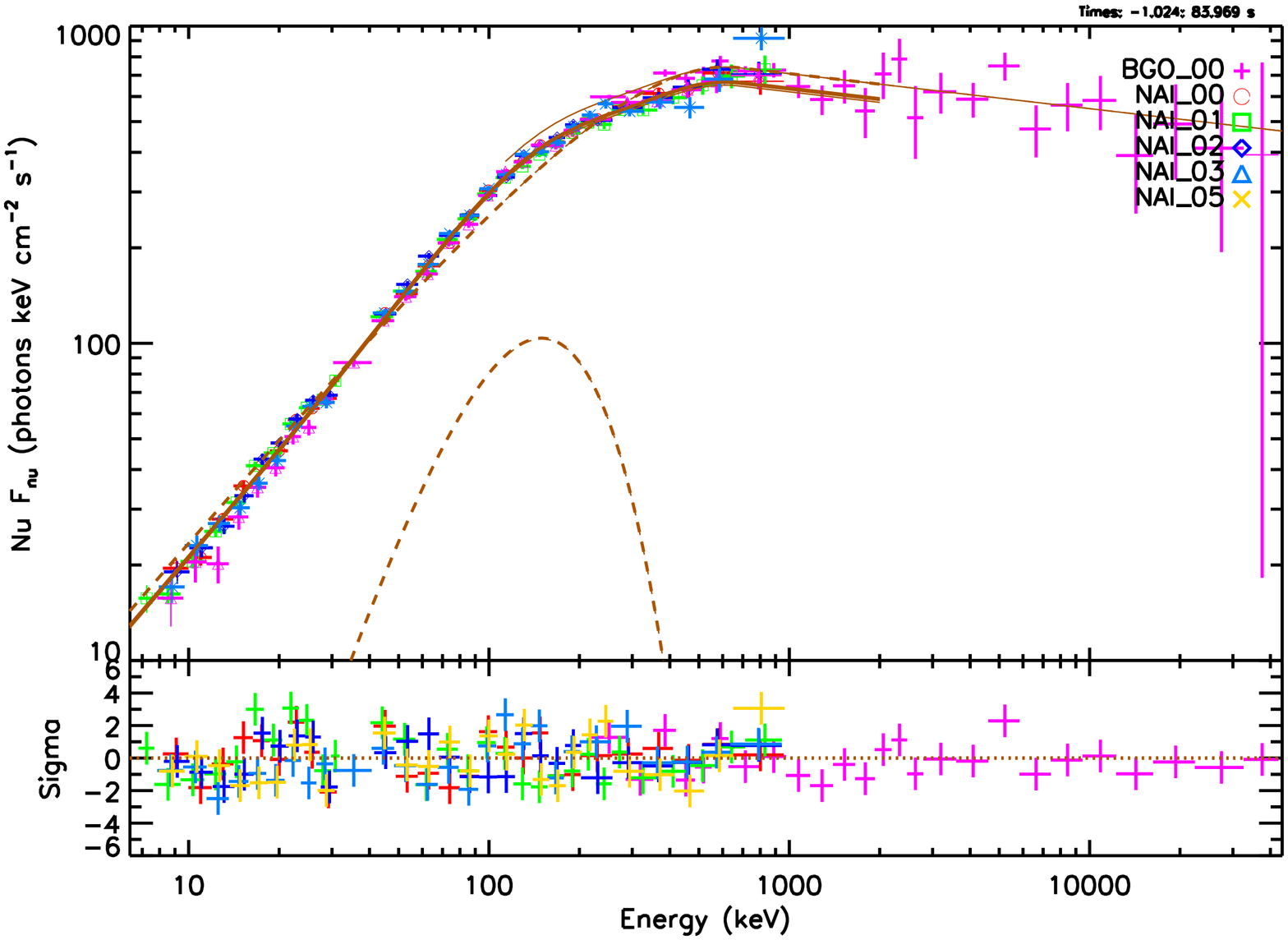}

\end{center}
\caption{The time-integrated spectrum of GRB~{\it 100724}B fit by a Band function (top two panels) and
a Band + blackbody function (bottom bottom two panels). The left plots show the count spectra for the two models and 
the right plots the corresponding deconvolved ${\nu}F_{\nu}$ spectra. The data points appear as color crosses. Dashed lines indicate the individual spectral functions and solid lines shows the summed model fit.
The addition of a BB spectral component over the brightest part of the
burst (T$_\mathrm{0}$-1.024s to T$_\mathrm{0}$+83.969s) shows a significant improvement in the fit
compared to a Band function by itself,
particularly noticeable as the removal of trends with energy in the residuals compared to the Band-only fit.
The region between 30 and 40 keV is excluded from the fit owing to
calibration issues around the k-edge of the NaI detectors.  We have verified this exclusion doesn't affect the
recovered parameter values.
\label{fig:figure2}}
\end{figure*}

We performed a time-integrated spectral analysis over the main part of the burst (T$_\mathrm{0}$-1.024s to T$_\mathrm{0}$+83.969s) using the Band function. The Band parameters are in part fairly typical of the ensemble of
GRBs, with $\alpha = -0.67 \pm 0.01$ and E$_{\rm peak}$ = $352 \pm 6$ keV \citep{Preece:2000,Kaneko:2006}. 
However, with an index $\beta = -1.99 \pm 0.02$, the high-energy power law systematically overshoots the observed flux above 1 MeV in BGO, as can be seen by the fit residuals in Figure~\ref{fig:figure2} (top two panels), which also indicate systematic patterns at low energy. This suggests a simple Band function does not adequately represent the spectrum of this burst.

We identify the best shape to fit the above-mentioned spectral deviation by fitting the same data simultaneously with a Band function combined with each of the following models: single power-
law (PL), Black Body (BB), Band function, power law with exponential cut off (``Comp" for Comptonized model), and Gaussian.
% and figures~\ref{fig:spectralDeviationFit1} and~\ref{fig:spectralDeviationFit2}). 
We select the best model by choosing the fit with the lowest Castor C-stat value (later C-stat). 
C-stat differs from the Poisson likelihood statistic by an offset which is a constant for a particular dataset.
Table~\ref{table:spectralDeviationFit} shows the results of these fits. The effective area correction described above is on the order of a few percent and does not change the C-stat for each fit more than a few units, nor does it change the value of the parameters resulting from the fit.

\begin{table*}[h!]
\caption{\label{table:spectralDeviationFit} Fit of the time-integrated spectrum of GRB~{\it 100724}B from T$_\mathrm{0}$+1.024s to T$_\mathrm{0}$+83.969s. The count spectrum using the NaI detectors 0, 1, 2, 3, 5 and BGO detector 0 is fit simultaneously with a standard Band function and with an additional model to evaluate the shape of the spectral deviation. Band+BB is preferred over all the other combinations.}

%\hspace{0.3cm}
%\begin{center}
%\rotatebox{90}{
{\footnotesize
\begin{tabular}{|l|l|l|l|l|l|l|l|l|l|l|l|l|l|}
\hline
\multicolumn{1}{|c|}{Models} &\multicolumn{3}{|c|}{Standard Model} & \multicolumn{10}{|c|}{Additional Model} \\
\hline
       &\multicolumn{3}{|c|}{Band} & \multicolumn{1}{|c|}{BB} & \multicolumn{2}{|c|}{Compt} & \multicolumn{3}{|c|}{Band} & \multicolumn{2}{|c|}{Gaussian} & \multicolumn{1}{|c|}{PL} & \multicolumn{1}{|c|}{Cstat/dof} \\
\hline
\multicolumn{1}{|c|}{Parameters} & \multicolumn{1}{|c|}{E$_{\rm peak}$} & \multicolumn{1}{|c|}{$\alpha$} & \multicolumn{1}{|c|}{$\beta$} & \multicolumn{1}{|c|}{kT} & \multicolumn{1}{|c|}{E$_{\rm peak}$} & \multicolumn{1}{|c|}{index} & \multicolumn{1}{|c|}{$E_{0}$} & \multicolumn{1}{|c|}{$\alpha$} & \multicolumn{1}{|c|}{$\beta$} & \multicolumn{1}{|c|}{Centroid} & \multicolumn{1}{|c|}{$log_{\mathrm{10}}$ FWHM} & \multicolumn{1}{|c|}{index} & \\
\hline
Band & $352$ & $-0.67$ & $-1.99$ & & & & & & & & & & $1133/704$ \\
     & $\pm6$ & $\pm0.01$ & $\pm0.01$ & & & & & & & & & &  \\
Band+BB    & $615$ & $-0.90$ & $-2.11$ & $38.14$ & & & & & & & & & $1038/702$ \\
           & $\pm29$ & $\pm0.02$ & $\pm0.02$ & $\pm0.87$ & & & & & & & & &  \\
Band+Compt & $708$ & $-0.94$ & $-2.13$ & & $164$ & $+0.81$ & & & & & & & $1039/701$ \\
           & $\pm48$ & $\pm0.02$ & $\pm0.02$ & & $\pm7$ & $\pm0.20$ & & & & & & &  \\
Band+Band  & $716$ & $-0.94$ & $-2.13$ & & & & $60$ & $0.76$ & $<-5$ & & & & $1039/700$ \\
           & $\pm48$ & $\pm0.02$ & $\pm0.02$ & & & & $\pm7$ & $\pm0.21$ & & & & &  \\
Band+Gaussian & $403$  & $-0.75$ & $-2.02$ & & & & & & & $103$ & $0.25$ & & $1060/701$ \\
           & $\pm8$  & $\pm0.01$ & $\pm0.01$ & & & & & & & $\pm2$ & $\pm0.03$ & &  \\
Band+PL    & $341$  & $-0.63$ & $-1.99$ & & & & & & & & & $-1.93$ & $1131/702$ \\
           & $\pm9$  & $\pm0.05$ & $\pm0.01$ & & & & & & & & & $\pm1.59$ &  \\
\hline
\end{tabular}
%}
}
%\end{center}
\end{table*}

 While spectral deviations from the standard Band function were previously
identified in the form of an additional PL to the Band function sometimes extending from the lower energy in the
 GBM to the higher energy in the LAT~\citep{Abdo:2009:GRB090902B,Ackermann:2010:GRB090510,Ackermann:2010:GRB090926,Guiriec:2010}, in the 
 case of GRB~{\it 100724}B a PL spectral component does not improve on the Band-only fit and an additional BB component to the Band function is the best model to fit 
 the spectral deviation. An equal C-stat is obtained for 
Band+Band and Band+Compt, but with $\alpha$ close to +1 for the additional Band and Compt functions, and a very low value for $\beta$ (only constrained as an 
 upper limit, below -5) for the extra Band function, the Band and Compt functions can be interpreted as a Planck 
 function. Even with more parameters, the additional Band and Compt functions resemble a BB component, reinforcing Band+BB as the best combination. Additional models were tried, such as a log-Parabola function 
 \citep{Massaro:2010}, but the results were highly disfavored, and we exclude them from Table 1.

 Figure~\ref{fig:figure2} (bottom two panels) shows the BB contribution below E$_{\rm peak}$. Compared to the Band-only fit, 
 E$_{\rm peak}$ is shifted towards higher energy to $615 \pm 29$ keV and $\beta$ is lower with a value of $-2.11 \pm 0.02$. This 
 index is consistent with the flux detected above 1 MeV, and the spectrum seen in the LAT~\citep{GRB100724B_GCN_LAT} at
higher energies. $\alpha$~is also significantly lowered to $-0.90 \pm 0.02$. 

While the simultaneous fit of all the selected detectors provides the best constraints on the two spectral components,
 fits with Band+BB to combinations of individual NaI detectors with BGO 0 result in similar parameter values and offer
significant improvement over the Band-only fit. This provides a check that the BB component is real and not
introduced by effects such as detector deadtime or spectral distortions that
would affect each detector in a different way depending on the angle of the
detector to the source.

To verify that the improvement in the fit obtained by adding a BB component to the Band function is not 
 a statistical fluctuation, we generated 20,000 synthetic spectra for each selected detector. For the simulations we used
 the parameters from the fit performed with the Band-only function, which we take
 as the null hypothesis. To create the simulated spectra, for each detector the real background is added to the source spectrum model 
 and  Poissonian fluctuations are applied to the sum. All the detectors are then fit simultaneously with both Band 
 and Band+BB, and their C-stat are compared.
None of the 20,000 simulated spectra give a difference larger than 45 units of C-stat ([Band]--[Band+BB]), while in the real data, this 
difference is 95 units, corresponding to a probability lower than $5 \times 10^{-5}$ that the BB excess is due to statistical 
fluctuations.

\begin{figure}[h!]
%\begin{center}
\hspace{-0.4cm}
\includegraphics[width=0.4\textheight]{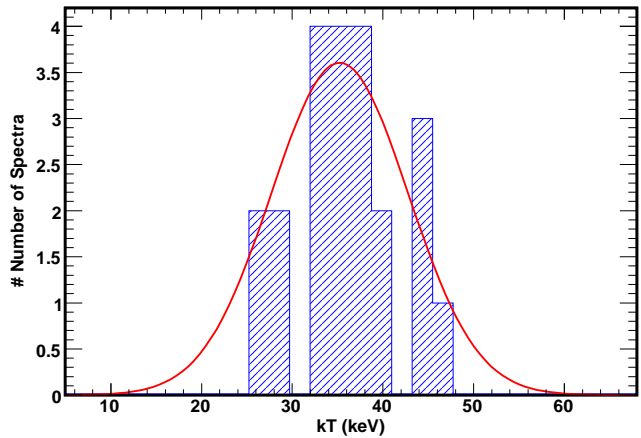}
%\end{center}
\caption{Distribution of the time-resolved black-body temperature kT. The spread in temperatures measured in the black-body component over the
22 time intervals shown in Figure~\ref{fig:figure1} can be fit by a Gaussian
distribution with a mean of $35.2 \pm 2.3$, and a $1\sigma$ standard deviation of $6.0 \pm 2.0$.
\label{fig:figure3}}
\end{figure}

To study the evolution of the spectral components, 22 time intervals were devised by requiring that each
interval produce a Band+BB spectral fit with well-constrained Band function
parameters, while attempting to separate the peaks and valleys of the light
curve so that the spectral fit parameters can be tracked with burst flux as
well as with time. The bottom panel of Figure~\ref{fig:figure1} exhibits the evolution of E$_{\rm peak}$ and BB temperature, kT, through 
these intervals. We notice that E$_{\rm peak}$ tracks the
lightcurve and globally decreases over time. The BB component is detected throughout the burst, and its temperature shows weak correlation with E$_{\rm peak}$. 
The significance of this correlation is difficult to assess, mostly because the variation in temperature is small. Overall, it appears the temperature is quite stable, with Figure~\ref{fig:figure3} showing more clearly the small scatter in kT.

\section{Discussion}

With a fluence of $\sim 5.2 \times 10^{-4}$ erg cm$^{-2}$
measured in 85 sec from T$_{0}$--1.024s between 8 keV and 40 MeV, GRB~{\it 100724}B is the most intense GRB detected by GBM over this energy range through 2010 September. Combined with the broad energy range of the GBM, this allows for accurate modeling of its energy spectrum even with this complex model.
Previous observational results regarding thermal components in GRBs were ambiguous 
and some were limited to individual fine time-slices rather than a spectral fit over the entire emission period. Some studies showing BB fits did not demonstrate that the BB fit was statistically preferred to a simple non-thermal component \citep{Ghirlanda:2003}. 
Other analyses found BB+PL spectra for isolated portions of selected GRBs, raising the 
possibility that these spectra are actually adequately fit with a standard Band function but that due to a weak signal in small time slices and extreme parameters for the Band function, a BB shape is competitive with the Band function \citep{Ryde:2004,Ryde:2005,Ryde:2010}. The non-thermal component fit with a single power law suggested a break well beyond
 the common E$_{\rm peak}$ values, and the BB temperature and its variations intriguingly matched those of a typical
 E$_{\rm peak}$. Despite the broader energy range of RHESSI GRB observations, one analysis found difficulties in fitting
 combined thermal plus non-thermal models~\citep{Bellm:2010}.

We find here that the joint BB plus non-thermal (Band)
fit is highly statistically preferred so that in simultaneously detecting both
components we are confident in their correct identification.
Time-resolved spectroscopy of GRB~{\it 100724}B reveals that this
BB component is seen throughout the burst and doesn't evolve much over time, while the non-thermal component follows the typical variations~\citep{Ford:1995,Guiriec:2010}. 
The consistency of the mean kT value with the temperature obtained in the time-integrated spectral fit,
combined with the detection of the BB component throughout the burst,
strengthen the case for an underlying thermal component in the gamma-ray
emission seen from GRB~{\it 100724}B and show that the presence of the BB in the time-integrated spectrum 
cannot be attributed to spectral evolution of the Band function during the burst.

E$_{\rm peak}$ varies substantially, from $\sim  90$ to $\sim 1100$ keV. At the same time, the thermal component remains relatively steady with the temperature varying only modestly between 30 and 50 keV as suggested by a $T\propto L^{1/2}$ dependence expected from a BB component. 
Time-averaged values of the temperature and the flux of the thermal component, and of the ratio of this flux over the total gamma-ray flux are 

$ kT=38 \pm 4 $ keV,  
$F_{\rm bb} = (2.6 \pm 1.4) \times 10^ {-7}$ erg/s/cm$^2$ and
$F_{\rm bb}/F_{\rm tot}= 0.04 \pm 0.02$.

In the standard fireball model these observables  allow determination of the 
physical properties of the outflow and its photosphere. 
Owing to the imprecise and delayed localization of GRB~{\it 100724}B, optical follow up to determine the distance to the source was impossible.
For this reason the temperature of the BB can be translated
into a real source temperature only as a function of source distance. 
We assume in the following argument a typical redshift $z=1$.
We find that the Lorentz factor is $\Gamma \simeq 325 \,  \xi^{1/4} f_\mathrm{NT}^{-1/4}$,
the photospheric radius is $R_{\rm ph}\simeq 5.6 \times 10^{11}\,\mathrm{cm} \,   \xi^{-3/4} f_\mathrm{NT}^{-1/4}$
and the radius at the base of the flow is $R_0 \simeq 1.2 \times 10^7\,\mathrm{cm}\, \xi^{-1} f_\mathrm{NT}^{3/2}$~\citep{Daigne:2002,Peer:2007}.
Here $\xi$ is a geometrical factor of order unity and $f_\mathrm{NT}$ is the efficiency of the mechanism responsible for the non-thermal emission. 
With an extreme efficiency $f_\mathrm{NT}=1$, these estimates are in good agreement with the typical values expected in the fireball model. The dependence on redshift is not strong: at z=3 (resp. 8), $\Gamma \simeq 645$ (resp. 1290), $R_{\rm ph}\simeq 1.1 \times 10^{12}\,\mathrm{cm}$ (resp. $1.4 \times 10^{12}\,\mathrm{cm}$), and $R_0 \simeq 1.1 \times 10^7\,\mathrm{cm}$ (resp. $6.9 \times 10^6\,\mathrm{cm}$).

Using more realistic values for the efficiency, 
the radius $R_0$ is the most altered, with $R_0\simeq \left(3.6-40\, \mathrm{km}\right)\xi^{-1}$ for $f_\mathrm{NT}\simeq 0.1-0.5$.
Such small values are puzzling. 
If the central engine is a rotating black hole, as in the popular collapsar model for long GRBs~\citep{Woosley:1993}, 
with a minimal mass in the range 5 -- 10 $M_\odot$, such radii are smaller than the typical value expected for the innermost stable orbit, from 44-89 km for a non-rotating black hole to 22-43 km for a highly-rotating black hole having a spin $a = 0.8$. 
These results for the time-integrated spectrum imply a small $R_{0}$ or a very large efficiency and the constraint is even stronger in some time bins.
We conclude that observations of GRB~{\it 100724}B require either a very high efficiency for the non-thermal process, or a very small size of the region at the base of the flow, both of
which are quite challenging for the standard fireball model, if not excluding it.

A simple solution to this
discrepancy between the standard fireball model and the observations is to assume
that the initial energy release by the central engine is not purely thermal,
but that the flow is highly magnetized close to the source~\citep{Daigne:2002,Zhang:2009}.
The magnetization
$\sigma$ is the ratio of the Poynting flux over the power (thermal + kinetic)
carried by the baryons. If no magnetic dissipation occurs below the
photosphere, the efficiency $f_\mathrm{NT}$ in the estimates of $\Gamma$,
$R_\mathrm{ph}$, and $R_0$ above should be replaced by
$(1+\sigma)f_\mathrm{NT}$. A magnetization $\sigma > 1$ will therefore reconcile the observations with physically acceptable values for the radius at the base of the flow and the efficiency of the mechanism responsible for the non-thermal emission. 
A similar conclusion is reached for scenarios where magnetic dissipation occurs early and contributes efficiently to the acceleration of the jet. However, the appearance of a low intensity thermal component in the spectrum probably excludes the most extreme version of the magnetized outflow scenario, where the energy is released by the central engine as a pure Poynting flux ($\sigma=\infty$).

%In conclusion, our results provide strong observational evidence for  the
%presence of a  photospheric spectral component, long suspected to exist in the
%standard fireball model. In addition, our results provide suggestive evidence
%for a significant magnetic component in the outflow.

\section{Conclusion}
We have shown that the simultaneous presence of thermal and non-thermal components
to the spectra of GRB~{\it 100724}B is statistically preferred.
Although the non-thermal component is dominant, the
black body flux is well within the GBM
sensitivity. Deviations from the Band function may be
measurable in less fluent bursts or in bursts where the thermal component is less prominent,
providing that the black body component lies in the band pass of the instrument and
its peak in energy is distinguishable from $E_{\rm peak}$. If the presence of an unresolved thermal component in other bursts modifies
the Band function parameters in the same sense as the Band-only fit for GRB~{\it 100724}B, then we might expect a systematic bias
yielding values of $\alpha$ and $\beta$ that are higher (harder) than in the true non-thermal component.
Two important consequences of this bias are that the perceived violation
of the synchrotron limit that disallows values $\alpha > -2/3$ (for slow-cooling electrons) and $\alpha > -3/2$ (for
fast-cooling electrons) may not be as common as suggested by~\citet{Preece:1998} and \citet{Crider:1997},
and that the relatively low rate of bursts detected by the LAT compared to the predictions
of~\citet{Band:2009} and the observations in Abdo et al. (in preparation)
based on extrapolations of $\beta$ from lower energies might be explained by this bias in $\beta$,
a possibility suggested also by~\citet{Ryde:2009}.

Our observations provide strong evidence for the presence of a photospheric spectral component, long suspected to exist in the
standard fireball model. In addition, our results require implausible parameters for the standard baryonic fireball model
and therefore favor a substantial magnetic component to the outflow.
 
\section{Acknowledgments}

%We thank the referee and the editor for their useful comments, which increased the quality of the paper.

The GBM project is supported by the German Bundesministerium f\"ur Wirtschaft und Technologie (BMWi) via the Deutsches Zentrum f\"ur Luft- und Raumfahrt (DLR) under the contract numbers 50 QV 0301 and 50 OG 0502.

A.J.v.d.H. was supported by an appointment to the NASA Postdoctoral Program at the MSFC, administered by Oak Ridge Associated Universities through a contract with NASA.

S.F. acknowledges the support of the Irish Research Council for Science, Engineering and Technology, cofunded by Marie Curie Actions under FP7.

P.M. acknowledges the support of NASA NNX08AL40G.

F.R. acknowledges the support of the Swedish National Space Board.

%References:
%done Meegan, C., et al. (2009) ApJ, 702, 791-804. 
%done \bibitem[Koshut~et~al.(1996)]{koshut96}Koshut, T.M., et al. 1996, \apj, 463, 570
%done \bibitem[Kouveliotou~et~al.(1993)]{kouv93}Kouveliotou, C., et al. 1993, \apj, 413, L101
%done \bibitem[Paciesas~et~al.(1999)]{paciesas99}Paciesas, W.S., et al. 1999, \apjs, 122, 465
%done \bibitem[Abdo~et~al.(2009)]....
%done W. S. Paciesas, M. S. Briggs, R. D. Preece and R. S. Mallozzi, in AIP Conf. Proc. 662, ed. G. R. %
% not find Ricker and R. K. Vanderspek, pp. 248--251 (2003)
%done G. Ghirlanda, G. Ghisellini and A. Celotti, A&A, 422, L55-L58 (2004)
%done G. Ghirland, L. Nava, G. Ghisellini, A. Celloti and C. Firmani, A&A, ???, ??? (2009)
%done E. P. Mazets, et al., in ASP Conf. Proc. 312, pp. 102--105 (2004).
%done L. A. Ford, et al., ApJ, 439: 307--321 (1995)

%\FloatBarrier
%\bibliographystyle{bibstyles/astron}
%\bibliography{2010-10-18_ApJL}

\end{document}